\documentclass{amsart}

\pdfoutput=1

\usepackage{graphicx}
\usepackage{subfig}
\usepackage{amssymb}
\usepackage{amsmath}

\begin{document}


\title[Dynamics of opinion in heterogeneous social networks]{Dynamics
  of bounded confidence 
opinion in heterogeneous social 
networks:\\ concord against partial antagonism}


\author[E. Kurmyshev]{Evguenii Kurmyshev}
\address[E. Kurmyshev]{Centro de Investigaciones en \'Optica \\ Loma del Bosque 115,
  Col. Lomas del Campestre, CP 37150 \\ Le\'on, Guanajuato, M\'exico}
\curraddr[E. Kurmyshev]{Centro Universitario de Los Lagos --- Universidad de
Guadalajara \\ Enrique D\'{\i}az de Le\'on 1144, Col. Paseos de la
Monta\~na, CP 47460 \\ Lagos de Moreno, Jalisco, M\'exico}
\email[E. Kurmyshev]{kev@cio.mx}
\email[E. Kurmyshev]{ekurmyshev@culagos.udg.mx}

\author[H.A. Ju\'arez]{H\'ector A. Ju\'arez}
\address[H.A. Ju\'arez \and R.A. Gonz\'alez-Silva]{Centro Universitario de Los Lagos --- Universidad de
Guadalajara \\ Enrique D\'{\i}az de Le\'on 1144, Col. Paseos de la
Monta\~na, CP 47460 \\ Lagos de Moreno, Jalisco, M\'exico}
\email[H.A. Ju\'arez]{hjuarez@culagos.udg.mx}
\thanks{H. A. Ju\'arez was supported by project CONACyT J50103-F}

\author[R.A. Gonz\'alez-Silva]{Ricardo A. Gonz\'alez-Silva}
\email[R.A. Gonz\'alez-Silva]{rgonzalez@culagos.udg.mx}
\thanks{R. A. Gonz\'alez-Silva was supported by project PROMEP/103.5/07/2636}


\keywords{Social networks, Opinion dynamics models, Relative
  agreement, Bounded confidence, Opinion bifurcation}


\begin{abstract}

Bounded confidence models of opinion dynamics in social networks have been
actively studied in recent years, in particular, opinion formation and extremism
propagation along with other aspects of social dynamics. In this work, after an
analysis of limitations of the Deffuant-Weisbuch (DW) bounded confidence,
relative agreement model, we propose 
the mixed model that takes into account two psychological types of individuals.
Concord agents (C-agents) are friendly people; they interact in a way that
their 
opinions get closer always. Agents of the other psychological type show partial 
antagonism in their interaction (PA-agents). Opinion dynamics in heterogeneous 
social groups, consisting of agents of the two types, was studied on different 
social networks: Erd¨os-R´enyi random graphs, small-world networks and complete 
graphs. Limit cases of the mixed model, pure C- and PA-societies, were also
studied.
We found that group opinion formation
is, qualitatively, almost independent of the topology of networks used
in this work. Opinion
fragmentation, polarization and consensus are observed in the mixed model at 
different proportions of PA- and C-agents, depending
on the value of initial opinion tolerance of agents. As for
the opinion formation and arising of ``dissidents'', the opinion dynamics
of the C-agents society was found to be similar to that of 
the DW model, except for the rate of
opinion convergence. Nevertheless, mixed societies showed dynamics and
bifurcation patterns notably different to those of the DW model. The influence
of biased initial conditions over opinion formation in heterogeneous social
groups was also studied versus the initial value of opinion uncertainty, 
varying the proportion of the PA- to C-agents. Bifurcation
diagrams showed impressive evolution of collective
opinion, in particular, radical change of left to right consensus or vice versa
at an opinion uncertainty value equal to 0.7 in the model with the PA/C
mixture of population near 50/50.
\end{abstract}




\maketitle


\section{Introduction}

Detailed behavior of every human being is the result of complex physiological
and psychological processes that are not well known yet. No one knows precisely
neither the dynamics of a single individual nor the way humans interact to each 
other. Therefore, any modeling of social dynamics inevitably involves a
significant simplification of a real problem. In modeling of social processes,
macroscopic phenomena are mainly due to the nontrivial collective effects
resulting from the interaction of a large number of ``simple'' elements of a
social network, rather than from a complex behavior of single entities. The way
to obtain useful results from this kind of modeling is to keep a balance between
the complexity of individual behavior and the complexity of a social network. 

The search for agreements and the reaching of consensus are important aspects of
social group dynamics, because they make the position of a group stronger and
enhance its impact onto society. Although the concept of opinion is not simple
to define formally (it can be quantified as a degree of satisfaction, desire or
preference), we see that opinion dynamics is an evolution problem, and 
therefore, it can be considered as a dynamical system. Examples are the
evolution of voting preferences and variations of market demand for products or
trademarks of competitors, among others. 

Many of the models recently proposed in social dynamics use techniques of
statistical physics~\cite{W71,CS73,G02,L81,SS00,DNAW00} (for a detailed review
of the topic and the state of the art see ~\cite{CFL07} and references in it).
The underlying concept of these models is a transition rate (or a probability
of 
transition) between different states of a social system, and opinion dynamics is
considered in terms of order-disorder transitions. 

Agent-based modeling on networks is another rapidly growing, deterministic
approach to social dynamics. Models of this kind study collective effects
resulting from the interaction of a number of ``simple'' agents in a social
network~\cite{S03}. Members of social groups are considered to be adaptive 
rather than rational agents, and no individual strategy is implied except a
common rule of opinion updating. Bounded confidence, relative agreement (BC/RA)
models of opinion dynamics are an important example  of agent-based 
modeling~\cite{AD04,DAWF02,D06,DNAW00,HK02,L07}.

Models of opinion dynamics are usually composed of the following basic elements:
opinion space, updating rule, updating dynamics and social network.

{\bf Opinion Space} Let  $S$ be an opinion space, such that every opinion $x_i$
of an agent $i$ is
in this space, $x_i \in S$. Two kinds of models are distinguished: models that
use a discrete opinion space, and others that consider a continuous opinion
space. In the first case, an opinion of an agent $i$ is usually a 
discrete-valued vector in $d$-Dim space, $x_i \in S \subset \{ 0, 1\}^d$, that 
represents the agent’s opinion over $d$ subjects (topics).  It should be noted 
that almost all computer simulations have been made for $d = 1$; in this case 
an agent has to choose between two options,  $\{ 0, 1\}$ or $\{ -1, 1\}$. Voter
model, majority rule and Ising spin models are examples supported by a discrete
opinion space~\cite{CFL07}. A unifying frame to incorporate all
discrete opinion dynamics models was proposed in~\cite{G05}.

There are situations, as the political orientation of a person, in which an 
agent preference changes smoothly within a range of possibilities, let us say 
from the extreme left to the extreme right. These situations are usually 
suitable to be treated within a continuous opinion space,
$x_i \in S \subset \mathbb{R}^d$. In practice, the bounded 1-Dim space is 
used, $x_i \in S \subset [ 0, 1]^d$ and $d=1$ ~\cite{L07}. 

{\bf Updating rule} All models assume that the change of opinion of an agent
depends on opinions of
agents related to him. The way to establish the relationship between agents in a
social network is commented later on. It depends on a particular social
network. 
When the relationship between agents is established, an agent takes into account
the opinions of other agents related to him if and only if those opinions are 
close to that of a given person. Closeness is usually defined by means of a
threshold that varies from a model to another. Some models assume that two
agents $i$ and $j$ interact with each other when $|| x_i - x_j || < \epsilon$,
where $\epsilon$ is a parameter of a model. Nevertheless, in other models a new
variable $u\in [0, 1]$, called opinion uncertainty or tolerance, is defined. 
Then, the closeness of interacting agents depends on $\epsilon$ and $u$. In both
cases, the updating rule assigns a new value to the opinion 
of a given agent $j$. This value depends on the value of its previous opinion 
and on the opinions of other agents close to the person.

{\bf Updating dynamics} Once the connections of an agent with others are
established, then we have to 
define the way they interact. It can be pair or group interaction. In the case 
of pair interaction, the latter is usually unidirectional. Among pairs of
connected 
agents, $(i, j)$, one agent is considered to be passive, say $j$, and the other 
to be active for a given time step. So, for this time step (time unit of 
updating) every passive agent $j$ updates his opinion as a function of the 
opinion of active agent $i$, not vice versa. In the case of group interaction, 
the opinion of passive (receptive) agent $j$ is updated in function of the 
average opinion of the agents connected to him. Once the updating of the opinion
of a certain number of agents has been carried out, we say that the one 
iteration or the time step of the model was done. The number of agents 
updating their opinions in one time step and the number of iterations depend on 
the model.

{\bf Social Network} Relationships between agents are usually described by means
of a network. A 
social network consists of a number of agents $N$, each represented by a vertex 
(node), and every pair of nodes of interacting agents is connected by an edge 
(link). The networks commonly used in computer simulation of social networks are
the complete network, the uncorrelated random graph proposed by 
Erd\"{o}s-R\'{e}nyi, the small-world network model by Watts and Strogatz, the 
growing networks (complex heterogeneous networks) by Barab\'{a}si and Albert, 
grids and real networks.    

\vspace{\baselineskip}

Most of the analysis of opinion dynamics focuses on a steady state 
opinion formation on a static or evolving social network~\cite{IKKB09}. The 
interest is in the study of possible opinion fragmentation, polarization or 
consensus in different social groups. The formation of steady state opinion 
clusters is interpreted as a locally ordered state of social groups in a
society,
in which the agents achieve a local consensus. In these groups, agents share 
ideas and common values, while a disordered state looks like a fragmented or 
anarchic society, in which it is impossible to reach agreements. 
	
In opinion dynamics, in particular in BC models,  an interaction between agents
is usually determined in the manner that the opinion of a passive agent tends 
to that of the active one. That is the case of pair interaction in the original
DW 
model~\cite{DAWF02}. In the Hegselman-Krause model, an agent adopts an 
average opinion of the nearest neighbors~\cite{HK02}. In these models, opinions
of 
interacting agents get closer to each other, and opposition (repulsive 
interaction) is not considered. Real life interaction between persons is 
repulsive-attractive usually. So, a number of mechanisms for repulsive 
interaction in BC models have been proposed and studied recently. 

In~\cite{JA05}, authors use a simple two-threshold, $U < T$, interpretation of
the Social Judgment Theory (SJT) in 1D attitude (opinion) space. The decision on
attractive or repulsive interaction is made according to the distance between 
the opinions of a randomly selected pair of agents, one of which is considered 
to be the passive one. When the distance $d_{12}$ between 
attitudes of a randomly selected pair of agents is less than $U$, the 
interaction is attractive; when $U < d_{12} < T$, then the attitude of the 
passive agent does not change; repulsion of opinions takes place when 
$d_{12} > T$. Thresholds $U$ and $T$ are free parameters of the model that are
used for all the population in experiments.  Neither opinion uncertainty no 
relative agreement are used in this approach.  
	
In papers~\cite{HDJ08,HD08} authors also refer to the SJT and propose 2D space
of attitudes $a_1$ and $a_2$. The second attitude $a_2$ is used as an 
indicator for triggering an attractive to a repulsive interaction in $a_1$. That
is the case of the asymmetric use of attitudes. The only function of $a_2$ is to
be a trigger to change an attractive into a repulsive interaction in $a_1$, and 
there is no direct influence of $a_1$ on $a_2$; there is no triggering in
$a_2$. 
The updating rule considers three different criteria based on the distances 
between the two attitudes of a pair of agents. One may expect that there has to 
be a direct mechanism of repulsive interaction for every given attitude, but not
only via an auxiliary attitude which is taken to be equally valid for all the 
population. Threshold $U = u_1 = u_2$ used in these works can hardly be 
interpreted as an uncertainty in the opinion of agents. So, neither opinion
uncertainty no relative agreement are used in this model,  as it should be in 
BC/RA models.
	
One of the most recent works~\cite{VPT10} considers the role of mass media and 
repulsive interactions in a BC continuous-opinion dynamics.  Authors introduce 
repulsive interaction in pairs of agents by random assigning positive and 
negative weights ($w_{ij} = w_{ji} = \pm 1$) to the links of a complete graph
(social 
network). Then a simple modification in the updating rule of 
the work~\cite{DNAW00} is used. Negative weight $w_{ij} = -1$ on the link
between 
agents $i$ and $j$ causes repulsive interaction between agents, while positive 
weight $w_{ij} = +1$ represents attractive interaction. So, when weights are 
assigned to all links of a network, the agents are divided into ``enemies'' and
``friends'' for further opinion evolution, because no links or weights are 
changed. This model takes into account the bound of confidence $\epsilon$, but
it does not use a relative agreement and opinion uncertainty. Another spin glass
model that takes into account this feature of "enemies" and "friends" was
developed 
in~\cite{G96}. A review of some Galam models that takes into account
heterogeneous beliefs, inflexible and contrarian effects can be found
in~\cite{G08}.

Others than BC approaches use repulsive interaction too. As an example, we refer
to the Axelrod models of social influence with cultural repulsion~\cite{RM10}, 
and to a continuous opinion dynamics model based on the principle of 
meta-contrast or Self Categorization Theory~\cite{S06}. 
	
In our study of opinion dynamics, we focus basically on social groups of 
adaptive rather than rational agents, no individual strategy is implied. We
study
the evolution of opinion of individuals, looking for the clustering of opinions.
A self-consistent mechanism of the repulsive-attractive interaction in the
frame 
of BC/RA continuous opinion model is proposed and studied. Randomness is a 
necessary feature of social interaction because both an individual attitude and 
the influence of social environment, are easily altered in time and space in a 
little predictable manner for an individual. Since bounded confidence models
are 
deterministic, stochastic features of opinion dynamics (social ``temperature'') 
are simulated in this work by means of seeding random initial distributions in 
opinion and tolerance, and personal links in social networks.   

The work is organized as follows. In Section~\ref{bcm}, we briefly describe and 
analyze the popular DW model of bounded confidence opinion dynamics. In
Section~\ref{cm}
we propose a new model of a mixed society that consists of PA- and C-agents. 
Results of computer simulation of the mixed PA/C-model on different social 
networks are presented in Section~\ref{ressim}.
Finally, conclusion and 
discussion are given in Section~\ref{conc}.


\section{Models}
\label{Mod-Prop}

\subsection{Bounded Confidence Models}
\label{bcm}

The \emph{Deffuant-Weisbuch} model, proposed in~\cite{DNAW00}, uses interval 
$S=[-1,1]$ as a continuous opinion space, so the opinion of agent $i$ is 
$x_i \in [-1,1]$. In addition, each agent $i$ is characterized by his opinion 
uncertainty $u_i\in (0, 1]$. 
An 
\emph{opinion segment} $s_i=[x_i - u_i, x_i +u_i]$ is assigned to each agent
$i$.
For two agents, $i$ and $j$,
their opinion segments overlap if and only if 
$h_{ij} = \min (x_i + u_i, x_j + u_j) - \max (x_i - u_i, x_j - u_j) > 0$, where 
$h_{ij}$ is called an opinion \emph{overlap}. If the overlap is strictly 
positive, then the \emph{relative agreement of agent $i$ with agent $j$} is 
defined by the following equation:  
\begin{equation}
\label{acuerdo}
\frac{h_{ij} - ( 2u_{i} - h_{ij})}{2u_i} = \frac{h_{ij}}{u_i} - 1.
\end{equation}

Opinion dynamics in a network is simulated as follows. Given a social network, 
a number of edges of the network (pairs of connected agents), $N$, is selected 
randomly, so that $N$ is usually equal to the number of agents in the network. 
In each pair, one of the interacting agents, say agent $j$, is sampled randomly
to be a \emph{passive (receptive) agent}, while the other one, say agent $i$, 
is considered to be the \emph{active agent}. If $h_{ij} > u_i$,  then the 
opinion and uncertainty of the passive agent $j$ are updated according to the 
following rule: 

\begin{equation}
\label{deffuant}
\begin{split}
x_j &= x_j + \mu \left( \frac{h_{ij}}{u_i} - 1 \right) (x_i - x_j) \\
u_j &= u_j + \mu \left( \frac{h_{ij}}{u_i} - 1 \right) (u_i - u_j)
\end{split}
\end{equation}
where  $\mu$ is a \emph{parameter of convergence} such that 
$\mu \in [0, \frac{1}{2}]$. If $h_{ij} \leq u_i$, there is no change in the 
opinion and uncertainty of agent $j$.

As one can see, the bounded confidence relative agreement DW model has the 
following distinctive features. During the interaction, agents change both 
their opinions and uncertainties. Agents with low uncertainty (high 
confidence) tend to be more influential than others. Equation~(\ref{acuerdo}) 
shows that relative agreement is linear in ${h_{ij}}/{u_i}$. Moreover, the 
condition $h_{ij} > u_i$, imposed in~\cite{DAWF02}, implies that the updating 
factor, Eq.~(\ref{acuerdo}) is always positive. As a consequence, both the 
opinion and uncertainty of passive agent get closer to those of the active 
agent, Eq.~(\ref{deffuant}). In other words, if there is interaction between two
agents, then the active agent convinces the passive one, and there are no 
possibilities of disagreement. The same initial value of uncertainty $U$ is used
for almost the whole population, adding a few agents with small values of 
uncertainty ~\cite{AD04}.

\subsection{Mixed PA/C-Model}
\label{cm}

Social groups are constituted of persons of different psychological types. 
In order to capture this important feature of social organization we propose 
a model of opinion in the mixed PA/C society. The PA/C-model involves agents 
of two psychological types, PA- and C-agents. The basic elements of the model, 
the opinion space, uncertainty and overlap of opinion intervals are treated in 
the same way as in the DW model. In the PA/C-model the opinion and uncertainty 
of passive (receptive) agent $j$ are changed when $0 \leq h_{i,j} \leq u_i$, in 
contrast to that of the DW 
model which excludes an interaction of agents at $h_{ij} < u_i$. Agents of the 
two types differ 
each other in the way they update their opinion in pair interaction.

The updating rule for the opinion and uncertainty of a passive C-agent $j$ is 
as follows: 
\begin{equation}
\label{concordia}
\begin{split}
x_j &= x_j + \mu \left( \frac{h_{ij}}{u_i} \right) (x_i - x_j) \\
u_j &= u_j + \mu \left( \frac{h_{ij}}{u_i} \right) (u_i - u_j).
\end{split}
\end{equation}
where active agent $i$ can be PA- or C-agent. Because the uncertainty of 
opinion is a qualitative rather than quantitative variable, we define a 
relative agreement for C-agents as $h_{ij}/u_{i}$, instead of 
$(h_{ij}/u_{i} - 1)$ in the DW model. When $h_{ij} < 0$, opinion segments do 
not overlap and therefore, there is no modification in the opinion and
uncertainty. 
The interaction of passive C-agents is always attractive in the opinion space,
similar 
to that in the original DW model. This behavior gives the name to the
psychological type, 
concord agents. There is no disagreement between interacting agents.

\begin{figure}
\centering
\includegraphics[width=0.4\textwidth]{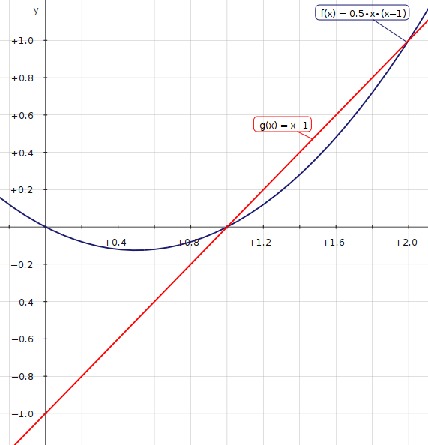}
\caption{The updating factor for PA-type agents.}
\label{scaling}
\end{figure}

Agents with partial antagonism (PA-agents) represent the other psychological
type. 
Note that in the original BC/RA DW model, the interaction between agents takes
place 
if and only if the overlap of opinion segments of connected agents is
sufficiently large, 
$h_{ij} > u_i$~\cite{DNAW00}. As a consequence of the imposed restriction, the
interaction 
between 
agents is always attractive, and the DW model does not allow any disagreement. 
Analyzing the interaction at smaller overlaps of opinion segments, $0 < h_{ij} <
u_i$, 
we see that is formally repulsive. Nevertheless, the repulsive regime is not
considered 
in the DW model. In addition, at this regime the updating factor is
mathematically 
symmetric to the attractive one (see Figure~\ref{scaling}). Real life
interaction 
between persons 
is usually repulsive-attractive. For that reason we modify the DW model,
breaking the 
symmetry of the updating factor in such a way that opinions of two interacting
agents 
can diverge, but not as strong as it could be in the DW model when the overlap
of opinion 
segments is less than an opinion uncertainty of the active agent. The new
updating rule 
for the opinion and uncertainty of a passive (receptive) PA-agent $j$ is as
follows:  
\begin{equation}
\label{antagonismo}
\begin{split}
x_j &= x_j + \mu_1 \left( \frac{h_{ij}}{2u_i} \right)\left(
  \frac{h_{ij}}{u_i} - 1 \right) (x_i - x_j) \\
u_j &= u_j + \mu_2 \left( \frac{h_{ij}}{2u_i} \right) \left(
  \frac{h_{ij}}{u_i} - 1 \right) (u_i - u_j).
\end{split}
\end{equation}
where active agent $i$ can be PA- or C-type. In contrast to the DW model, the
scaled 
relative agreement $(h_{ij}/2u_i) (h_{ij}/u_i - 1)$ is used for PA-agents. The
scaling 
factor 
$(h_{ij}/2u_i)$ decreases repulsion of opinions since $(h_{ij}/2u_i) (h_{ij}/u_i
- 1) 
\in [-0.125, 1]$ 
for PA-agents, and $(h_{ij}/u_i - 1) \in [-1, 1]$ (see Figure~\ref{scaling}). 
In addition, we relax 
the restriction $h_{ij} > u_i$  of the DW model to the condition $h_{ij} > 0$,
allowing 
negative values for the relative agreement $(h_{ij}/2u_i) (h_{ij}/u_i - 1)$
when 
$0 < h_{ij} < u_i$. There is no modification of opinions and uncertainties when 
$h_{ij} < 0$.
Two different convergent parameters $\mu_1$ and $\mu_2$  are used, because the
dynamics 
of opinion is qualitatively similar to the dynamics of uncertainty, but the 
updating rates of these variables can be certainly different.

The PA/C-model considers a society of $N$ individuals of two different
psychological 
types: a fraction $p$ of agents are of the C-type, while the rest of population
are 
PA-agents.
The society consists of two 
sub-populations, $M_C$ and $M_{PA}$ of sizes $pN$ and $(1-p) N$ respectively, 
being $0 \leq p \leq 1$ a parameter of the model. When an active agent $i$
interacts with a passive one $j$, agent $j$ changes its opinion and uncertainty
following the rules defined for C-agents if $j\in M_C$ or those
defined for PA-agents if $j\in M_{PA}$.

Varying the value of $p$ from $0$ to $1$, we model mixed societies with 
populations ranging through C- and PA-type societies.
In order to get a more 
realistic simulation, we use heterogeneous initial conditions instead of the 
homogeneous distribution in initial opinion that was used in~\cite{DNAW00,D06}. 
Homogeneous initial opinion in a social network is obtained by drawing 
individual opinions from the interval $[-1, 1]$ with the uniform probability 
distribution. In this work, we define a parameter $0 \leq \ell \leq 1$ that 
describes ``political preference'' of agents. The initial opinions of $\ell N$ 
agents are drawn from the interval $[-1, 0]$ with the uniform probability 
distribution (these agents are considered to be left oriented). Opinions of the
rest of population, $(1 - \ell)N$ right oriented agents, are drawn from the 
interval $[0, 1]$ with the uniform probability distribution also. When 
$\ell = 0.5$, we have homogeneous initial conditions. So, the new model is 
doubly heterogeneous:
\begin{itemize}
\item Parameter $p$ regulates the psychological composition of the society 
\item Parameter $\ell$ changes the initial preference of agents, so we can 
model right (or left) oriented societies. 
\end{itemize}
These two characteristics result in a rich dynamics of the model.


\section{Computer simulation results}
\label{ressim}

Computer simulations of the model
were carried 
out in a way slightly different to that in~\cite{DNAW00,D06}.
The state 
variables of a social network are the opinion $x_i$ and the uncertainty $u_i$ of
individuals. 
Uniform probability distribution is used to choose the
initial uncertainty of each agent from the interval $[U - 0.2, U + 0.2]$, where
$U$ is the average initial opinion uncertainty in the network. 
This condition differs of that in~\cite{DNAW00,D06},
where all agents use the
same initial value $U$ for uncertainty. $U$ is a parameter of the model, and we
vary it in the interval $0.3 \leq U \leq 1.2$.

At each time step, we choose randomly $N$ pairs of coupled agents, edges
of a social network. Then, one agent of each pair is selected at random to be
the active agent $i$, while the other is considered to be the passive one, $j$.
The value of the overlap $h_{ij}$ is computed, and we update variables of the 
passive agent if and only if $h_{ij}\geq 0$.
As mentioned above, the composition of heterogeneous society
is regulated with the parameter  
$0 \leq p \leq 1$, 
so that $M_C = pN$ and $M_{PA} = (1-p) N$ is the number of C-
and PA-agents, respectively. When an active agent $i$ interacts with a 
passive one $j$, agent $j$ changes its attitude according to the
C-type rules if 
$j\in M_C$, or the PA-type rules if $j\in M_{PA}$. 

 The value of the parameter $U$ runs through $0.3$ to $1.2$ 
at a step of $0.01$. At each value of $U$, we execute $350$ iterations (time 
steps). The results shown in the following figures represent an average over 
$50$ simulations for each value of $U$. Each simulation starts with a new seed 
of initial conditions in opinion distribution and uncertainty. 

Three kinds of networks are used in our experiments. The first is the complete 
network with $400$ agents. The second one is a random network of $1000$ agents, 
such that the degree $d_i$ of each vertex $i$ satisfies $30 \leq d_i \leq 40$, 
and the distribution of the degrees over the network is uniform. The third 
network studied in this work is a Watts-Strogatz small world network of $1000$ 
agents, with $k = 20$ and $\beta= 0.25$.

First of all, we are interested in the study of homogeneous societies with
primitive democracy. So, neither hubs nor leaders are considered and the
initial opinion of each agent is a number randomly drawn from interval 
$[-1, 1]$, using uniform probability distribution, $\ell = 0.5$.
Because we are basically interested in the
steady state of opinion dynamics, we plot the density of asymptotic
opinions as a function of parameter $U$. In the following figures, axis 
$x$ is opinion, and the vertical axis is the number of agents that share an 
opinion at a given value of $U$. We observed that all bifurcations patterns
of opinion distributions
obtained in this work are qualitatively similar for the three social networks 
under consideration. So, here after, we show only the results 
obtained for a Small World Network.

\begin{figure}
\centering
\subfloat[]{
\label{mixto1:a}
\includegraphics[width=0.42\textwidth]{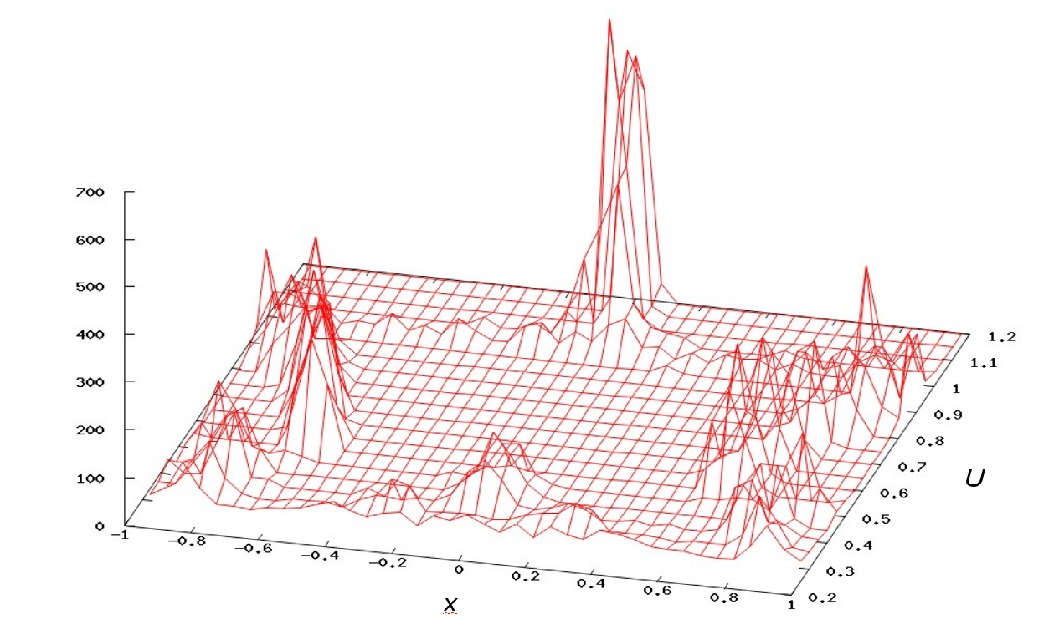}}
\hfill
\subfloat[]{
\label{mixto1:b}
\includegraphics[width=0.42\textwidth]{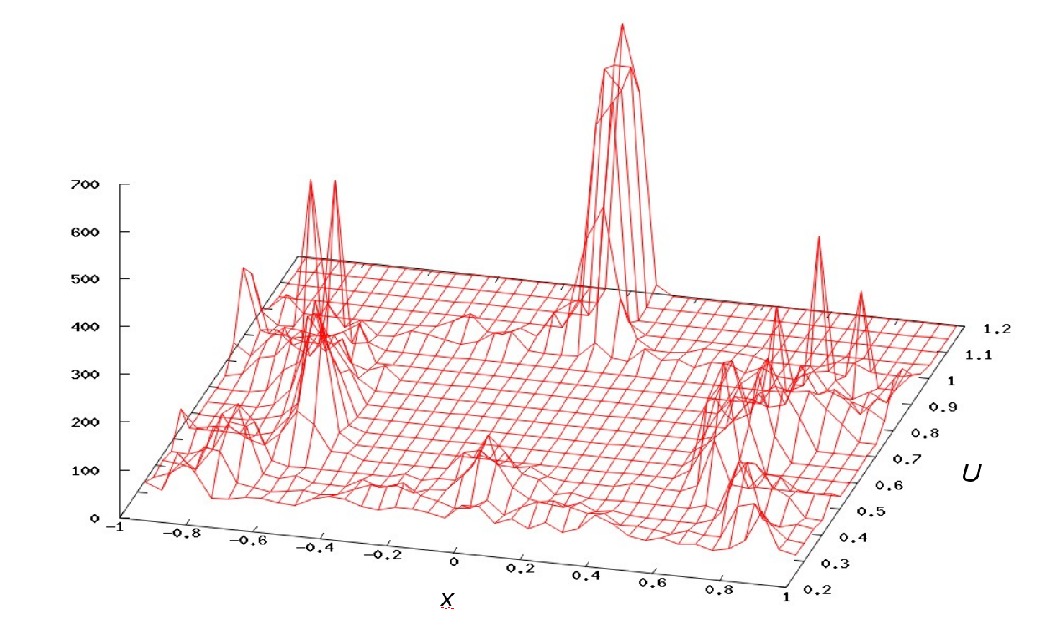}}
\\[\baselineskip]
\subfloat[]{
\label{mixto1:c}
\includegraphics[width=0.42\textwidth]{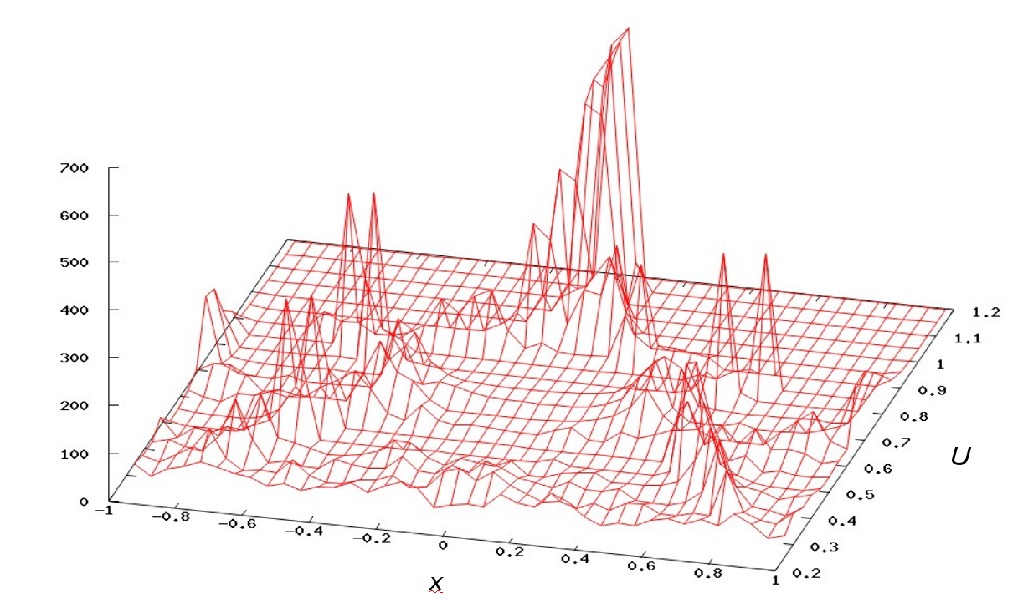}}
\hfill
\subfloat[]{
\label{mixto1:d}
\includegraphics[width=0.42\textwidth]{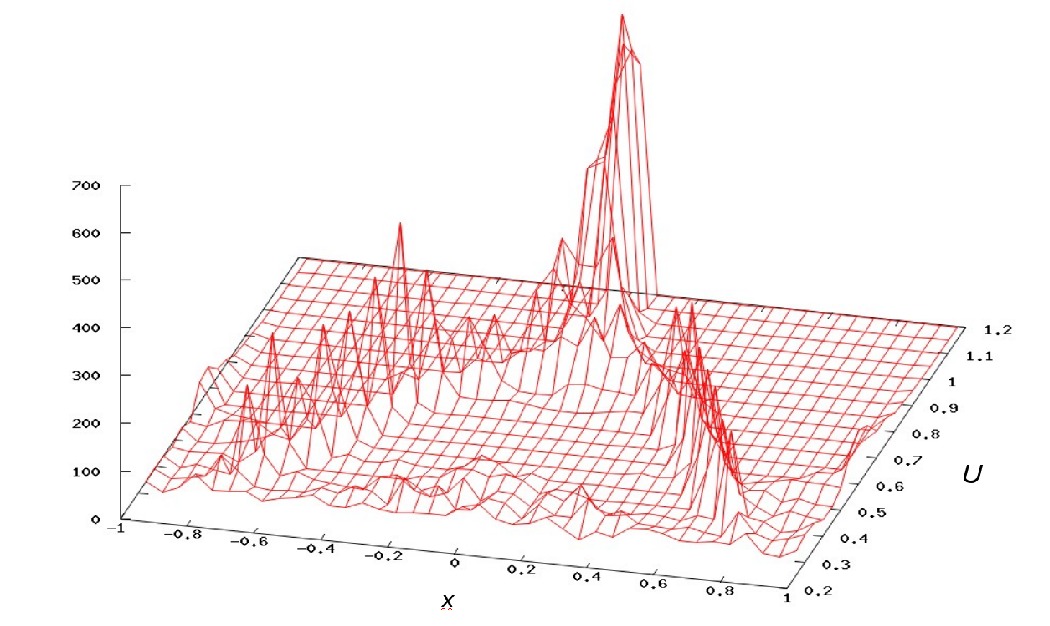}}
\\[\baselineskip]
\subfloat[]{
\label{mixto1:e}
\includegraphics[width=0.42\textwidth]{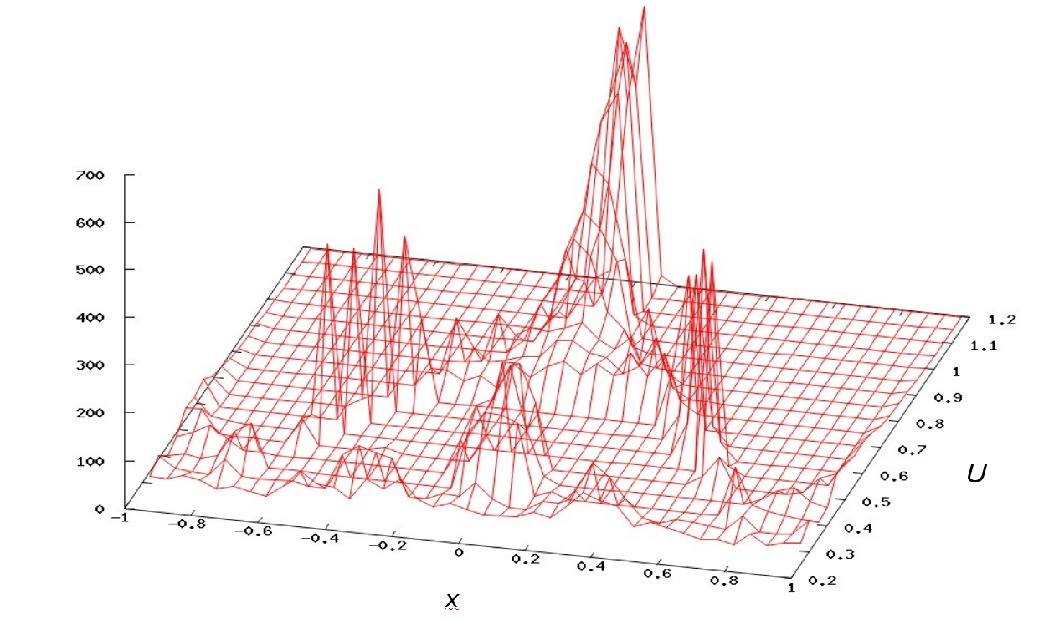}}
\hfill
\subfloat[]{
\label{mixto1:f}
\includegraphics[width=0.42\textwidth]{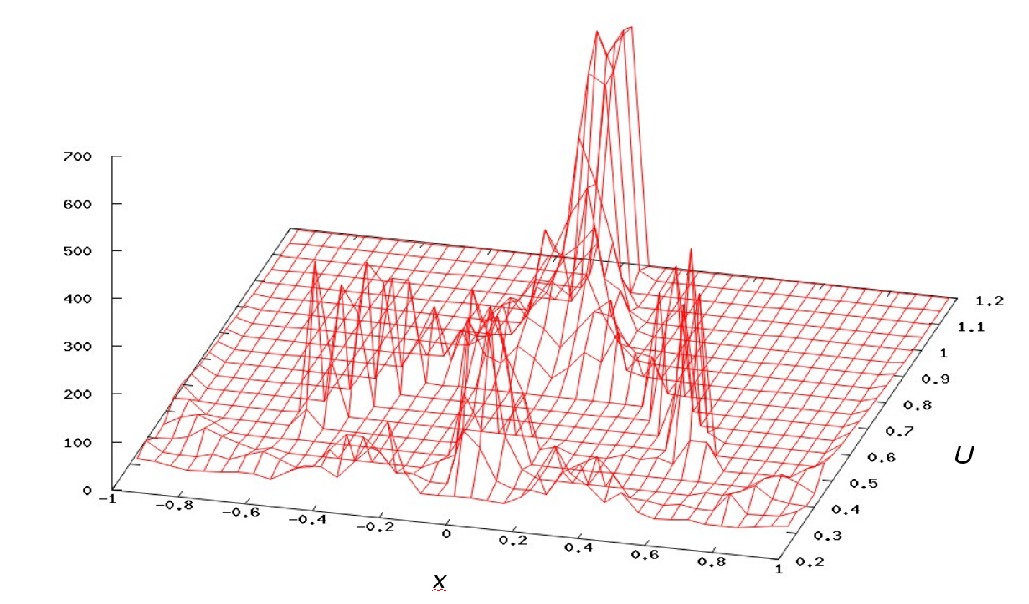}}
\\[\baselineskip]
\subfloat[]{
\label{mixto1:g}
\includegraphics[width=0.42\textwidth]{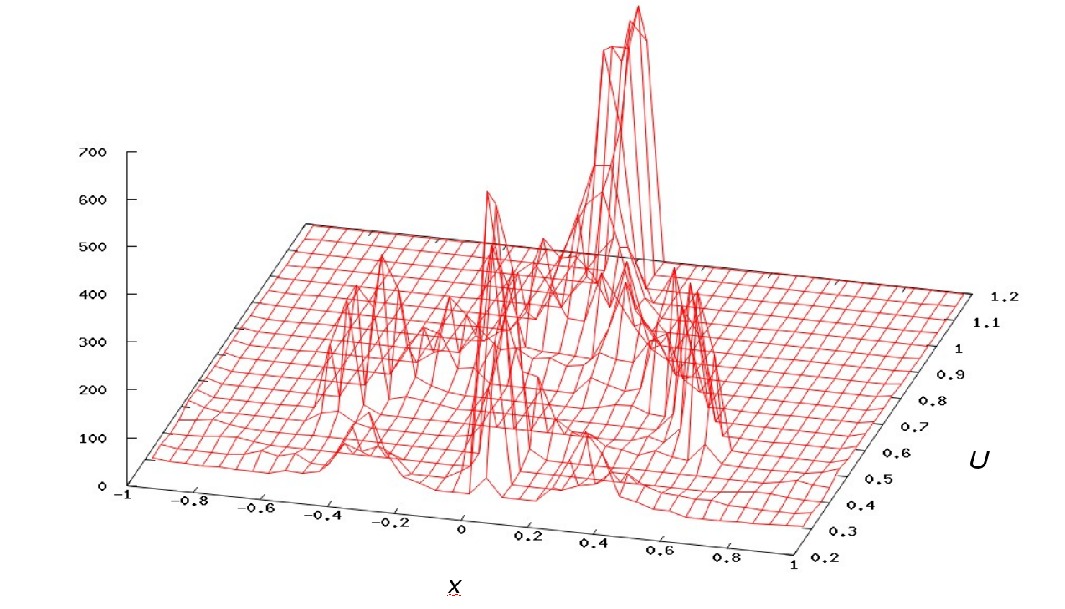}}
\hfill
\subfloat[]{
\label{mixto1:h}
\includegraphics[width=0.42\textwidth]{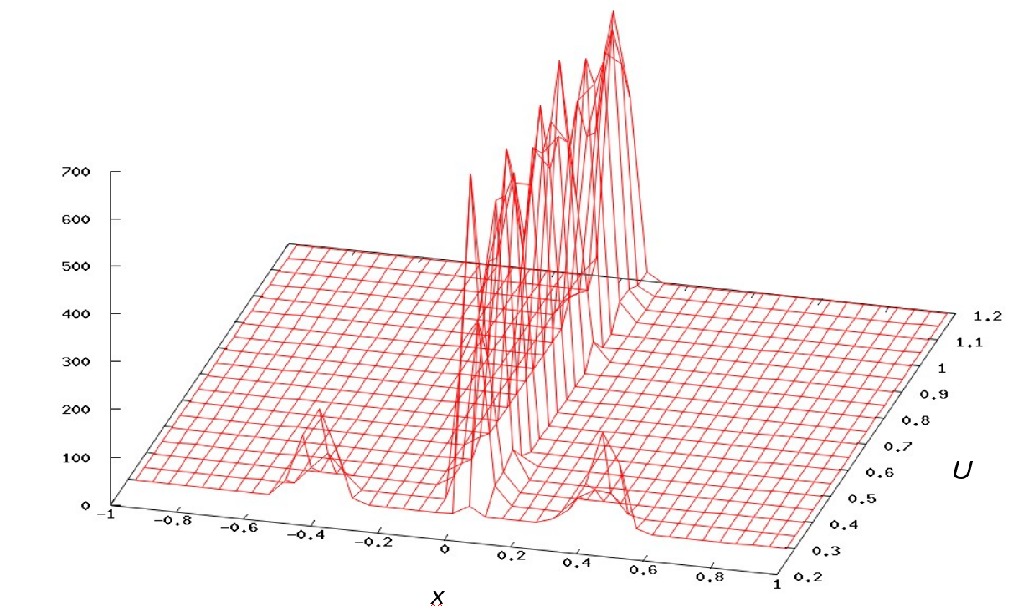}}
\caption{Bifurcation of opinion distribution in the mixed model for different 
values of $p$. 
Initial opinions were drawn from the interval $[-1, 1]$, (a) $p = 0.0$, 
(b) $p = 0.1$, (c) $p   = 0.25$, (d) $p = 0.4$, (e) $p = 0.6$, (f) $p = 0.75$, 
(g) $p = 0.9$   and (h) $p = 1$.}
\label{mixto1}
\end{figure}

Figure~\ref{mixto1}
shows bifurcation patterns (an average over $50$ simulations) of group opinions
for different values of $p$, ranging from $p = 0$ to $p = 1$, and
uniform distribution of initial opinions of  
agents in the interval $[-1, 1]$. Uniform initial distribution of uncertainties
in the interval $[U - 0.2, U + 0.2]$ is used for each value of $U$.
 
For $p = 0$, the society is composed of PA-agents.
Bifurcation of opinion distribution as a function of 
parameter $U$ is shown in Figure~\ref{mixto1:a} in the interval 
$0.3 < U< 1.2$. 
Four opinion clusters are 
observed for $U < 0.4$, although they are not well separated. For 
$0.4 \leq U \leq 0.55$ there are three well defined opinion groups. In the 
interval $0.55 \leq U \leq 1$, there are two opinion groups. Note that the two 
opinions get more distant from each other as $U$ increases from $0.55$ to $1$; 
in other words, as the initial opinion uncertainty (tolerance) of agents gets 
larger, the society polarizes in two almost extreme opinion
groups. However, at $U = 1$ there is another phase transition, and the
two opinions collapse into a unique opinion, consensus. So, 
three bifurcation points are observed at $U = 0.4, 0.55, 1$. 


For $p = 1$, the society is composed of C-agents. Bifurcation of opinion 
distribution is shown
in Figure~\ref{mixto1:h}.
Note that for $U \leq 0.4$, the C-society is divided in three groups of 
agents, two of them with opposite opinions. For $U > 0.4$ there is a unique 
opinion, consensus, being $U = 0.4$ a bifurcation  point of the system. Results 
of simulations show that the opinion dynamics in the C-model is qualitatively 
independent of the networks used.

At intermediate values of $p$, a series of 
transformations of the bifurcation pattern is observed (see Fig.~\ref{mixto1}, 
b through g).  For $0 < p < 0.25$ and $U > 0.45$, the bifurcation of the two 
opinion clusters goes to the consensus through a collapse (see Fig.~\ref{mixto1} 
(a,b and c)), in 
contrast to the “soft convergence” of the two opinion clusters into the 
consensus for $0.25 < p < 0.9$ (see Fig.~\ref{mixto1}, d through g). 
Fig.~\ref{mixto1} (e,f and g) shows that a mixed society reaches 
a kind of consensus in the interval $0.4 < U < 0.5$, which becomes 
unstable for $U > 0.5$,
splitting into two group opinions. Then, the two opinions merge softly into 
the consensus for $U > 1$. For $p \approx 1$, a mixed society reaches a 
consensus at $U > 0.35$.


Another interesting feature of opinion dynamics in a mixed society is observed 
in the series of Figs.~\ref{mixto1} (e, f, g). Figure~\ref{mixto1:g} shows 
opinion dynamics of the C-population diluted with 10\% of PA-agents. The 
bifurcation pattern in Figure~\ref{mixto1:g} is similar to that in 
Fig.~\ref{mixto1:h} for the pure C-society, in the interval of values 
$0.3 < U < 0.5$. Nevertheless, the consensus breaks out (splits) into two 
opposite opinions ($\langle u \rangle = + 0.6, -0.6$) at $U = 0.6$. Then, 
those two opinions get closer as $U$ increases, merging into the consensus at 
$U = 0.85$, once again. Similar behavior of opinion is observed in 
Figs.~\ref{mixto1} (e,f). The important thing is that a relatively small 
fraction of PA-agents changes opinion dynamics significantly.


Opinion dynamics in a mixed society becomes even more interesting when we 
“switch on” non-uniformity in the initial opinion distribution. Let 
$\ell \in [0, 1]$. We randomly select $\ell N$ agents and assign them initial 
opinions randomly drawn from the interval $[-1, 0]$ with uniform probability 
distribution; those are left oriented agents. For the rest of agents,
$(1 - \ell)N$, we draw their initial opinions from the interval $[0, 1]$
with uniform probability; they are right oriented. So, the parameter $\ell$ 
controls the initial amount of individuals that share the same orientation in 
their opinions.

\begin{figure}
\centering
\subfloat[]{
\label{mixto000:a}
\includegraphics[width=0.48\textwidth]{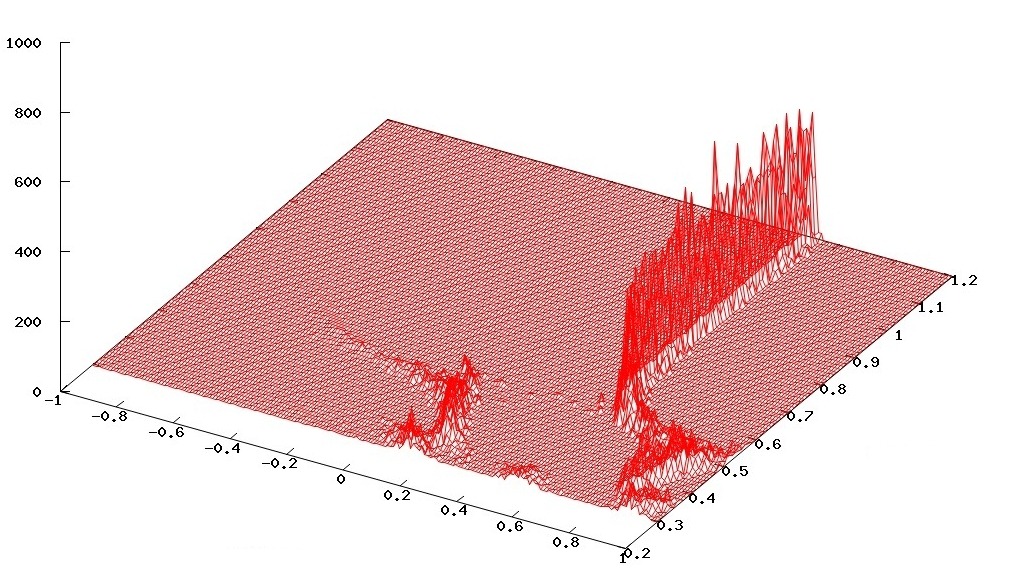}}
\hfill
\subfloat[]{
\label{mixto000:b}
\includegraphics[width=0.48\textwidth]{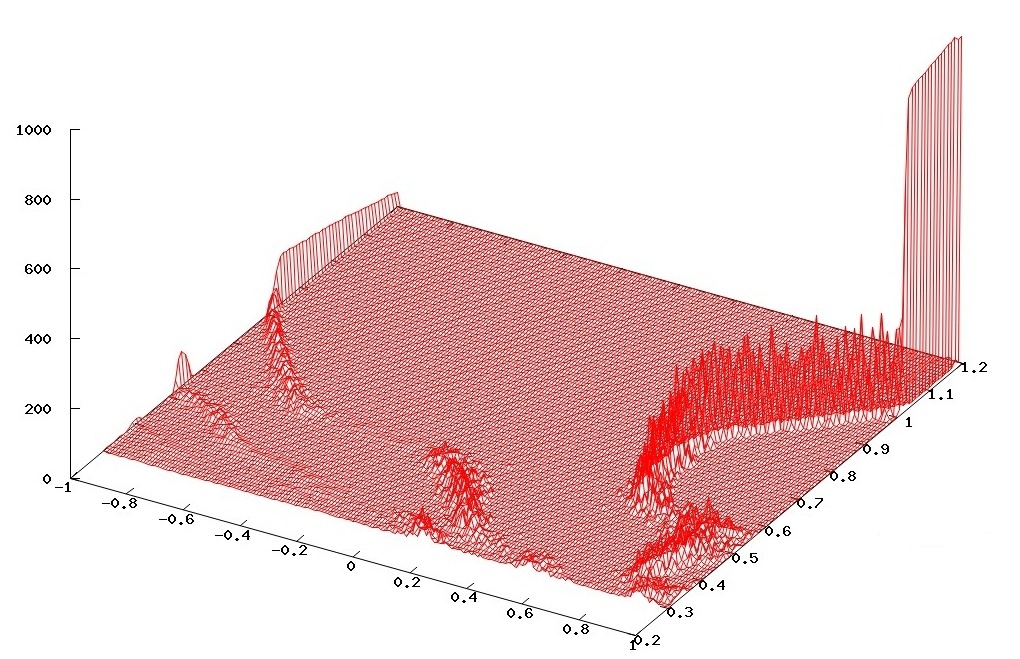}}
\\[\baselineskip]
\subfloat[]{
\label{mixto000:c}
\includegraphics[width=0.48\textwidth]{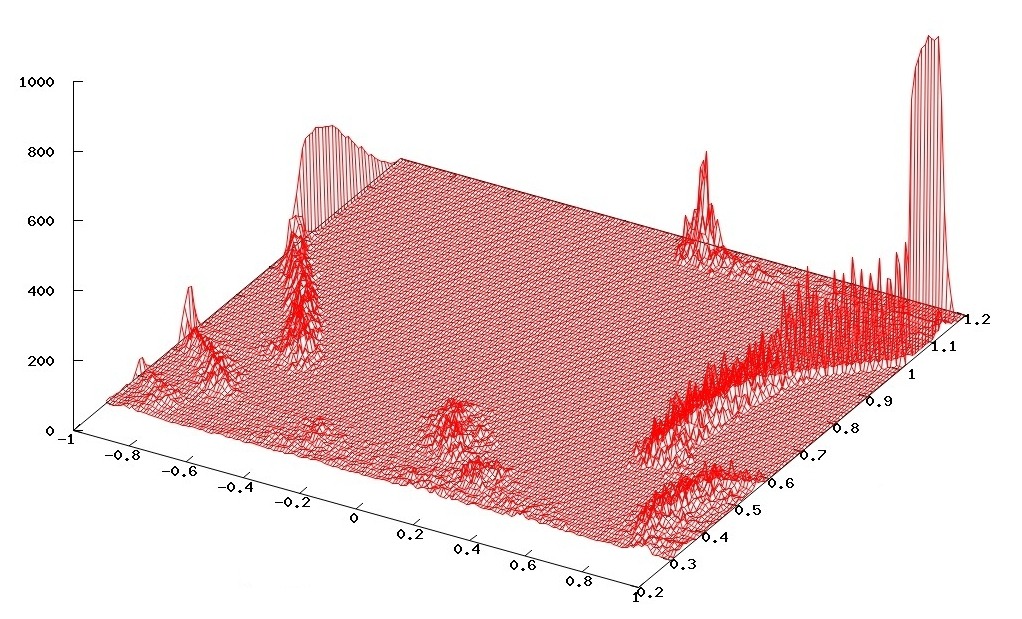}}
\hfill
\subfloat[]{
\label{mixto000:d}
\includegraphics[width=0.48\textwidth]{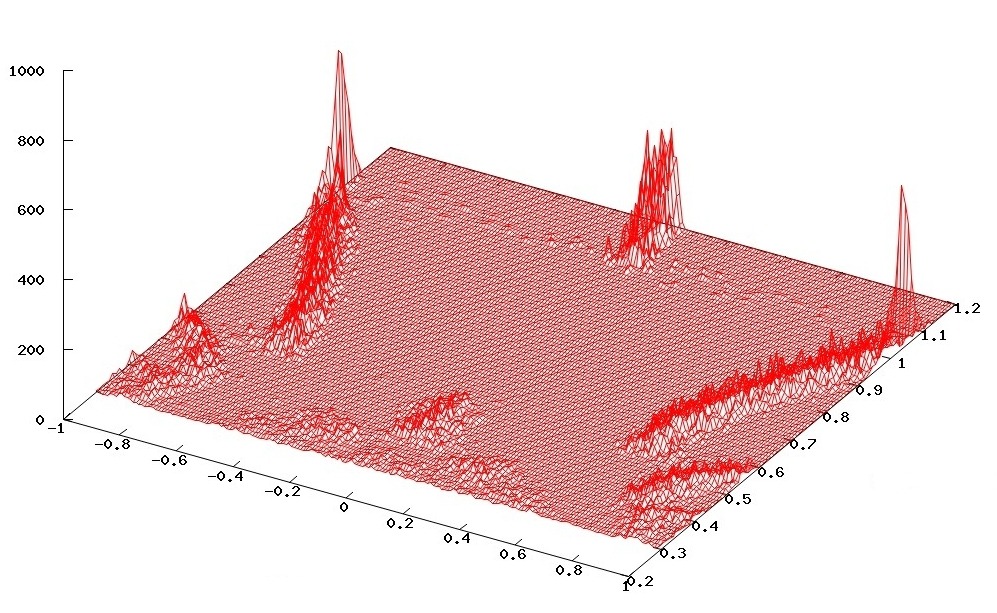}}
\caption{Bifurcation of opinion distribution as a function of $U$ in the mixed
  model for $p = 0$ and (a) 
$\ell = 0.0$, (b) $\ell = 0.2$, (c) $\ell   = 0.4$, (d) $\ell = 0.5$.
}
\label{mixto000}
\end{figure}

We run a number of simulations for different values of $p$ and $\ell$. Results 
of simulation have shown that for a given value of $p$, the graphs of opinion 
corresponding to $\ell$ and $1 - \ell$ are symmetric, as expected. So, we show 
results only for the right oriented societies ($\ell \in [0,0.5]$). 
Figure~\ref{mixto000} shows results of computer simulation for $p=0$ 
(a society of PA agents) and 
different values of $\ell$. 
The bifurcation pattern for a totally right oriented society, $\ell = 0$, is 
shown in 
Fig.~\ref{mixto000:a}. This pattern is similar to the picture in 
Fig~\ref{mixto1:a}
that was slightly modified and scaled to the interval
of opinions $[0, 1]$. We see the formation of the unique opinion cluster 
centered at $\langle x \rangle = 0.5$ for $0.55 \leq U$. However, there is a
small cluster of left oriented agents that do not approve the ``main'' 
opinion. As $\ell$ increases, this group of ``dissidents'' also grows, and the 
society gets polarized, as shown in Figures~\ref{mixto000:b} 
and~\ref{mixto000:c}. In these figures, for $U > 0.55$ there are two clusters of
agents with almost opposite opinions, cluster of right oriented agents
being greater than the other. 
The size of the small cluster increases as $\ell$ increases. As $U$ increases, 
opinions of the two groups diverge tending to the extremes. The two extreme
opinions collapse into the consensus at $U = 1$, for $\ell > 0.4$. 
For $\ell = 0.5$, we come back
to the PA-model with the uniform initial distribution of opinions (compare 
Figure~\ref{mixto000:d} with Figure~\ref{mixto1:a}).


For a fraction of C-agents $p < 0.35$, computer simulation showed that the 
bifurcation pattern of opinion dynamics does change little compared to this of 
Figure~\ref{mixto000}. So, in the societies composed mainly of PA agents, 
opinion dynamics remains basically the same.


\begin{figure}
\centering
\subfloat[]{
\label{mixto050:a}
\includegraphics[width=0.48\textwidth]{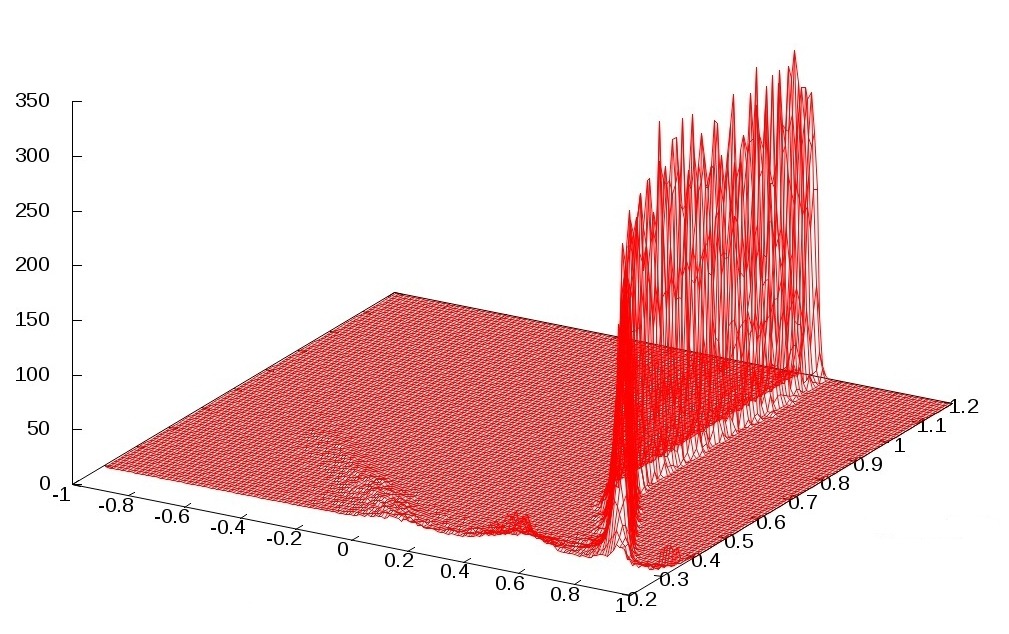}}
\hfill
\subfloat[]{
\label{mixto050:b}
\includegraphics[width=0.48\textwidth]{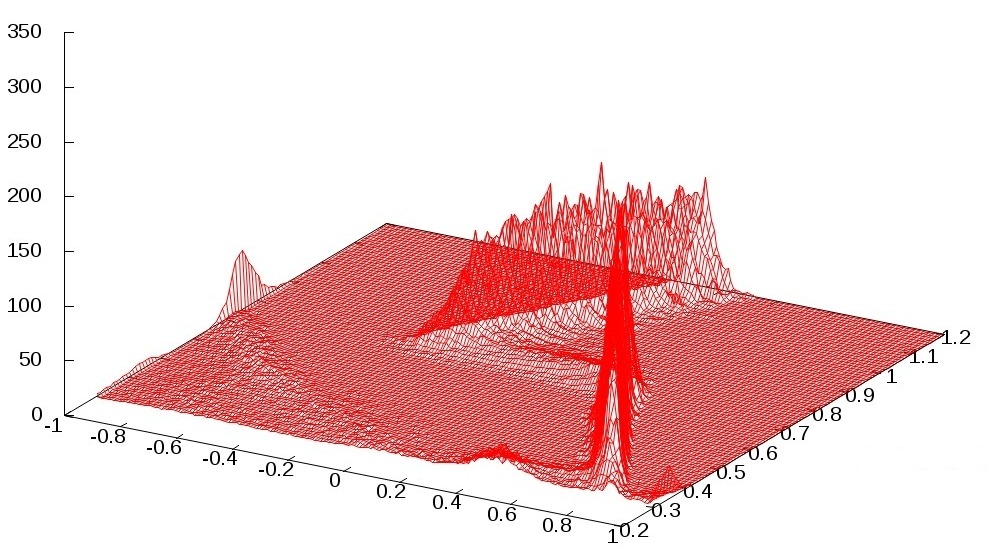}}
\\[\baselineskip]
\subfloat[]{
\label{mixto050:c}
\includegraphics[width=0.48\textwidth]{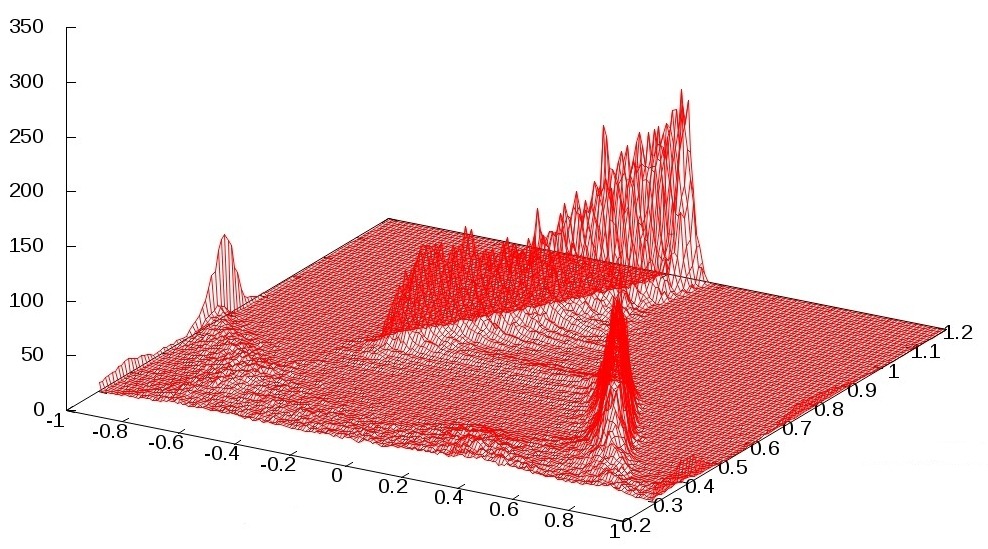}}
\hfill
\subfloat[]{
\label{mixto050:d}
\includegraphics[width=0.48\textwidth]{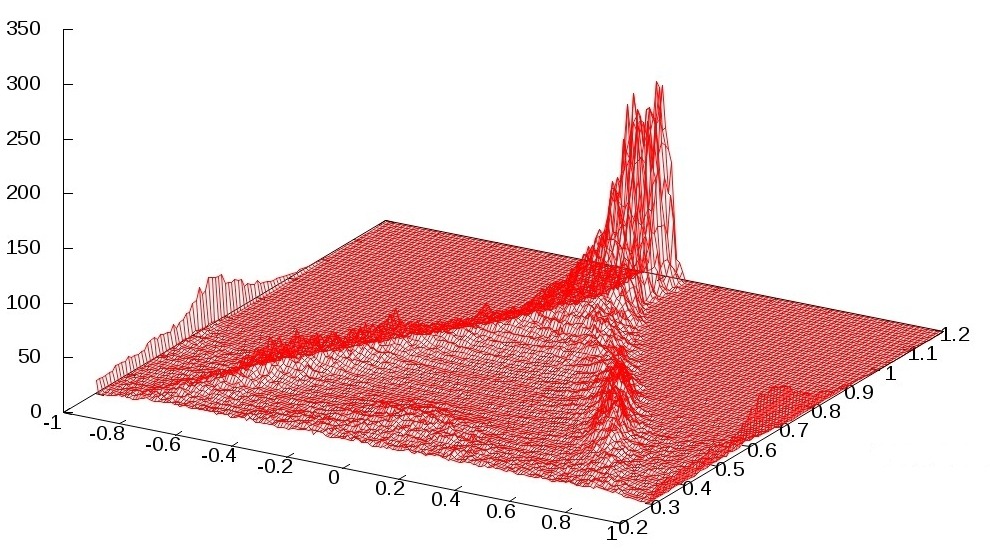}}
\caption{Bifurcation of opinion distibution as a function of $U$ in the mixed model at 
$p = 0.50$ (half of the agents are C-type agents, half are PA-type agents) and 
(a) $\ell = 0.0$, (b) $\ell = 0.2$, (c) $\ell   = 0.4$, (d) $\ell = 0.5$.
}
\label{mixto050}
\end{figure}

\begin{figure}
\centering
\subfloat[]{
\label{mixto050:det:a}
\includegraphics[width=0.48\textwidth]{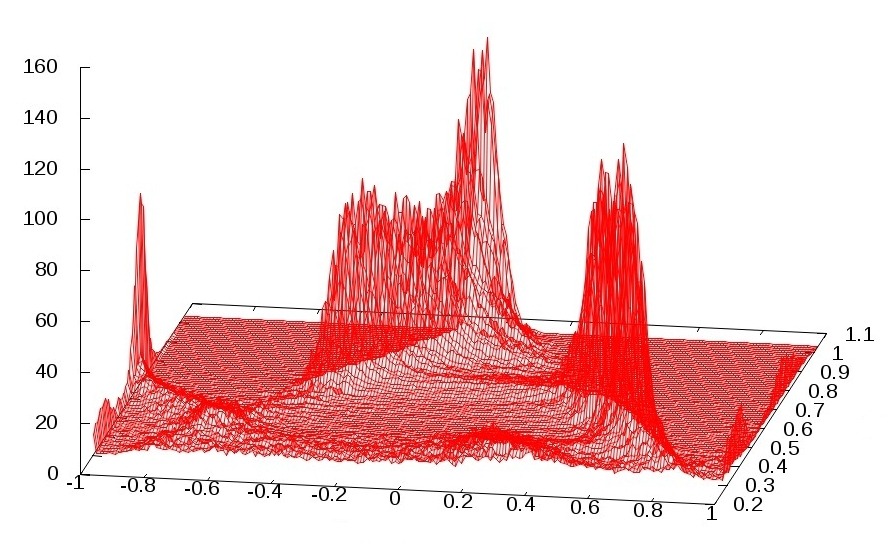}}
\hfill
\subfloat[]{
\label{mixto050:det:b}
\includegraphics[width=0.48\textwidth]{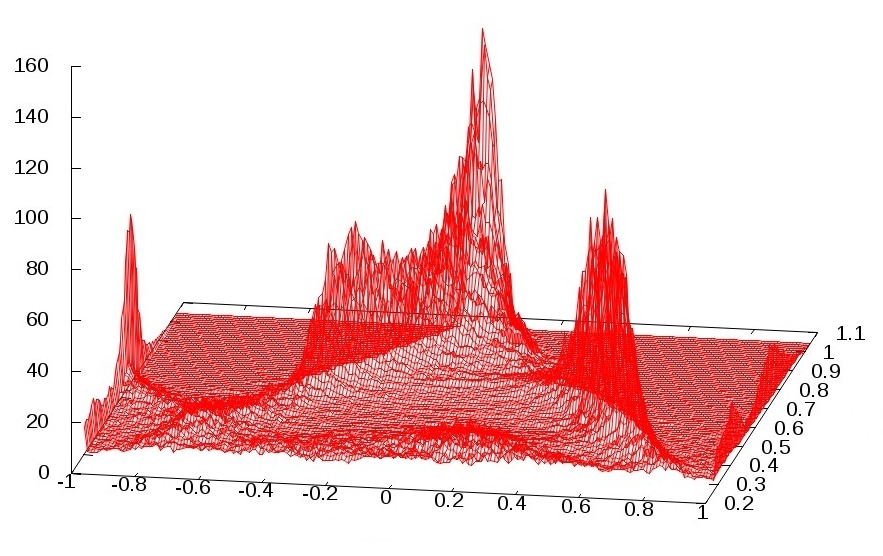}}
\\[\baselineskip]
\subfloat[]{
\label{mixto050:det:c}
\includegraphics[width=0.48\textwidth]{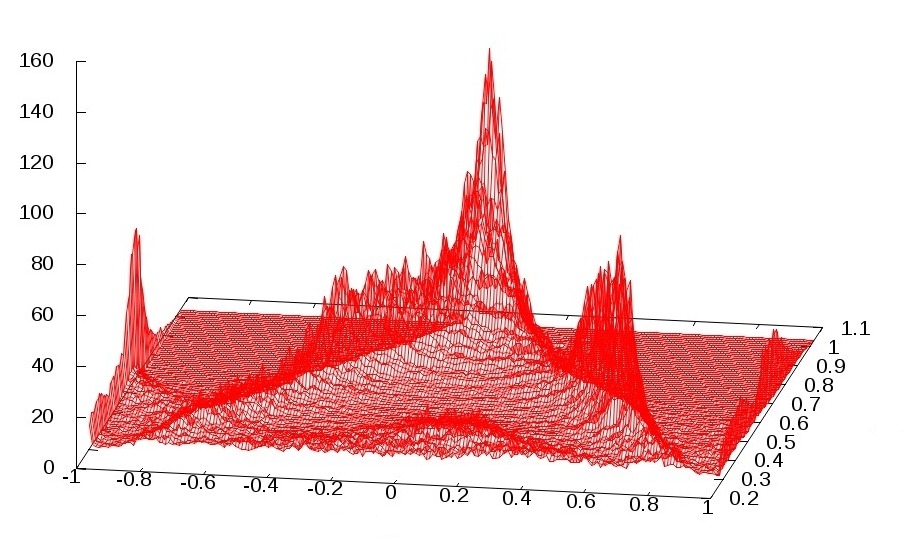}}
\caption{Bifurcation of opinion distribution as a function of $U$ in the mixed model at 
$p = 0.50$ (half of the agents are C-type agents, half are PA-type agents) 
and (a) $\ell = 0.42$, (b) $\ell = 0.45$, (c) $\ell   = 0.47$.
}
\label{mixto050:det}
\end{figure}

Significant changes were observed in opinion dynamics for $p > 0.35$. 
Figure~\ref{mixto050} shows bifurcation patterns for $p=0.5$, when one half of 
agents are PA-agents, and the other half are C-agents. In 
this case the opinion dynamics of the society is strongly influenced by the 
C-component of the population. For $\ell = 0$ the only one, right-oriented 
opinion dominates in the society; at the very beginning of the interval of 
uncertainties $0.3 < U < 1.2$, we observe two small groups of agents with 
different opinions to the majority of the population. These small groups 
disappear at $U > 0.4$. When the fraction of left-oriented agents changes in 
the interval $0 < \ell < 0.45$, a steady state opinion dynamics reveals a 
little expected behavior of the society (Figures~\ref{mixto050:b} 
and~\ref{mixto050:c}). For $0.45 < U < 0.7$ there are two opinion groups, one 
big group of agents on the right and a small cluster of dissidents on the 
left. For $0.6 < U < 1$ there is a unique group of opinions, but the dominant 
public opinion suddenly changes from the right to the left at $U = 0.7$. Thus, 
at $U = 0.7$ the public opinion suffers a phase transition from right to left 
oriented one. This is a 
surprising result, showing that the steady state opinion of a society can be 
very sensitive to the value of tolerance $U$. For $1 < U$, the whole 
population endorses an opinion close to $0$, and the society becomes center 
oriented. Figure~\ref{mixto050:d} shows the steady state opinion 
dynamics when $p = 0.5$ and $\ell = 0.5$, which is similar to 
Fig.~\ref{mixto1:d}. Two opinion clusters get closer to each other as $U$ 
increases, and then merge into one cluster, resulting in a $\lambda$ type 
pattern. We see that the dynamics shown in Figure~\ref{mixto050:c}, 
$\ell = 0.4$, and Figure~\ref{mixto050:d}, $\ell = 0.5$, are qualitatively 
different. Figure~\ref{mixto050:det} shows the transition between the two 
dynamics. In Figure~\ref{mixto050:det:a} ($\ell = 0.42$), we observe a dynamics
similar to that in Figure~\ref{mixto050:c}, which changes smoothly from 
Figure~\ref{mixto050:det:b} ($\ell = 0.45$) to Figure~\ref{mixto050:det:c} 
($\ell = 0.47$), finally resulting in Figure~\ref{mixto050:d}.


\section{Conclusion and Discussion}
\label{conc}

In this work, we proposed a model of 
opinion formation in 
a heterogeneous society
consisting of agents of two psychological types, with concord 
and partial antagonism behavior. Clustering of agents in opinion space was 
studied. To this end, we proposed a
bounded confidence, relative agreement 
model with agents updating their opinions by means of one of the two
rules.
A 
concord agent (C-agent) always gets its opinion closer to that of another
agent, 
this differs from the DW model in the way we define relative
agreement. PA-agent gets its opinion
farther from or closer to that of another  
agent depending on their relative agreement. In terms of physics, this means a 
repulsive-attractive potential between agents in the opinion space. Opinion 
formation in a society of agents of different psychological types was simulated
in the mixed model varying the ratio of PA- to C-agents. 
	
In order to study opinion formation in time, pair interaction between agents 
was used in the model. Varying the initial mean value $U$ of the opinion 
tolerance of agents, in our model we observed fragmentation, polarization or 
consensus. Computer simulations show that a steady state opinion is 
qualitatively little sensitive to the three networks used in the proposed model. 
In addition, a smaller opinion tolerance 
(uncertainty) causes a larger fragmentation of opinion. So, a society of close 
minded persons tends to be partitioned into a large number of small groups of 
agents with similar opinion. Details of fragmentation of a society in the 
opinion space depend on the updating rule of opinion and a social network.

The dynamics of opinion formation and bifurcation patterns  
depend on the value of the parameter $p$ in the mixed model.
For example, time convergence to one of the 
asymptotic opinion states (fragmentation, polarization or consensus) was much 
faster in the case of C-agents ($p = 1$) than that of the
mixed models ($0 \leq p <1$). As we 
expected, opinion formation in the C-society
was qualitatively similar to that of
the DW model, even though the direct comparison of the results is not possible
because we used initial conditions for opinion and uncertainty (tolerance) 
different to those in~\cite{DNAW00}. We found that 
in the mixed model the 
transition of opinion from one asymptotic state to another is a bifurcation, 
depending on the initial mean value of opinion uncertainty of agents. Even 
though bifurcations are observed in all bounded confidence models studied 
in~\cite{DNAW00,L07}, these differ from the bifurcations observed in our model,
especially for $p < 1$. The main difference is that the DW model
shows repetitive interruption of the line of the centrist opinion cluster 
followed by scaling in the bifurcation diagram (see Fig. 1 in~\cite{L07}), while
branching in our model for $p < 1$ goes from odd to even number of 
$1-2-3-4-\ldots$ branches without scaling (see Fig.~\ref{mixto1}). 
All the lateral branches of the bifurcation diagram in the 
DW model tend to converge to the central line (centrist opinion) as the initial
tolerance $U$ increases, in contrast to our model where lateral branches 
of fragmented opinion
clearly diverge for $p < 1$ as $U$ increases. In addition, an interesting 
phenomenon was observed in the bifurcation pattern for $p = 0$ and uniform 
initial distribution of opinion (see Fig.~\ref{mixto1:a}); when initial 
uncertainty in opinion increases in the interval $0.55 \leq U \leq 1$, two 
equal groups of agents have opposite opinions that diverge almost to the 
extremes, $-1$ and $1$, until they suddenly collapse into the consensus. It
looks 
like the more open-minded social groups tend to separate more from each other 
before they reach a consensus. Our vague ``sociological'' explanation of this 
dynamics is that ``clever = open-minded'' groups of agents initially tend to 
emphasize their differences in opinion (idea). However, when they become as 
``clever = open-minded'' as they could understand and accept the opponents’ 
idea, they get to the consensus. This bifurcation pattern differs from that of 
the DW model, in which the two branches of polarized opinion converge 
gradually into the consensus~\cite{DNAW00,L07}. In this concern, the
sociologists’ interpretation of the observed dynamics would be very much 
valuable. A formal mathematical analysis, classification and comparison of 
bifurcations of all the bounded confidence opinion models should be done. 

An important feature of this work is the use of biased initial conditions in 
opinion besides the uniform distribution of initial opinion in a social group, 
in contrast to what is usually considered in previous works~\cite{DNAW00,L07}, 
with the exception of maybe a particular case of a society of open- and 
close-minded agents~\cite{L10}. Uniform initial conditions can be interpreted as
a state of complete democracy that further evolves to a symmetric fragmentation
of a steady state opinion, usually. But how do opinions evolve when a social 
group initially has two subgroups of different size, and the average opinion of
one subgroup is ``left-oriented'' and that of the other is ``right-oriented''?
To simulate this situation, we used a piece-wise uniform distribution of initial
opinion, varying the ratio of a number of left- to right-oriented agents, 
including limit cases when one of these subgroups does not exist. The simulation
of opinion formation in our mixed society shows an extremely 
interesting behavior of the society. When nearly half the population of a social
group were PA-agents, and the other part were C-agents, $p \approx 0.5$, we 
varied the ratio of carriers of ``left'' to ``right'' ideas, 
parameter $\ell$, from $0$ to $1$. When $\ell$ was in the interval 
$\ell = (0.15, 0.42)$ (the society initially has ``right'' ideas), we found that
at $U = 0.7$ , the main branch of a steady state opinion diagram bifurcates 
drastically from ``right'' to ``left'' (the opinion of the majority of the 
group changes from $+0.5$ to $-0.5$, approximately) , or vice versa, when 
$\ell = (0.58, 0.95)$. From a sociological point of view, it is a critical 
behavior in opinion formation. This ``mechanism'' can explain unexpected 
results in a voting process, when the social composition and the initial opinion
state of a society have not been assessed correctly. In addition, when the main
branch is asymmetric, subgroups of opponents (dissidents) and centrists are also
observed. A tiny dissidents’ branch that tends to the opposite extreme opinion 
when the initial uncertainty increases was always observed for any tolerance, at
$p = 0.5$. This can provide a mechanism for the formation of extremism.  

As it has been shown, the main features of opinion formation as fragmentation,
polarization, consensus, centrism and extremism in opinion space emerged 
naturally in our models. In addition, the Mixed model studied in this work is 
``doubly'' heterogeneous. First, it describes a heterogeneous society with a 
different ratio of PA- to C-agents. Second, we use heterogeneous initial 
conditions in opinion and tolerance. Piece-wise homogeneous distributions are 
used for initial opinion. Also, we assign different initial tolerances to 
agents within a relatively wide interval of values near the mean value $U$, so 
that a society has a variety of agents between close- and open-minded ones. All 
these characteristics show that the model proposed and studied in this work
provide a mechanism by means of which the formation 
of opinion in different social groups can be simulated and explained.



\bibliographystyle{amsalpha}
\bibliography{Socialbib}

\providecommand{\bysame}{\leavevmode\hbox to3em{\hrulefill}\thinspace}
\providecommand{\MR}{\relax\ifhmode\unskip\space\fi MR }
\providecommand{\MRhref}[2]{%
  \href{http://www.ams.org/mathscinet-getitem?mr=#1}{#2}
}
\providecommand{\href}[2]{#2}
\begin{thebibliography}{DNAW00}

\bibitem[AD04]{AD04}
Fr\'ed\'eric Amblard and Guillaume Deffuant, \emph{The role of network topology
  on extremism propagation with the relative agreement opinion dynamics},
  Physica A \textbf{343} (2004), 725--738.

\bibitem[CFL07]{CFL07}
Claudio Castellano, Santo Fortunato, and Vitorio Loreto, \emph{Statistical
  physics of social dynamics}, Submitted to Reviews of Modern Physics, 2007.

\bibitem[CS73]{CS73}
Peter Clifford and Aidan Sudbury, \emph{A model for spatial conflict},
  Biometrika \textbf{60} (1973), no.~3, 581--588.

\bibitem[DAWF02]{DAWF02}
Guillaume Deffuant, Fr\'ed\'eric Amblard, G\'erard Weisbuch, and Thierry Faure,
  \emph{How can extremism prevail? a study based on the relative agreement
  interaction model}, Journal of Artificial Societies and Social Simulation
  \textbf{5} (2002), no.~4.

\bibitem[Def06]{D06}
Guillaume Deffuant, \emph{Comparing extremism propagation patterns in
  continuous opinion models}, Journal of Artificial Societies and Social
  Simulation \textbf{9} (2006), no.~3, 8.

\bibitem[DNAW00]{DNAW00}
Guillaume Deffuant, David\ Neau, Frederic Amblard, and G\'erard Weisbuch,
  \emph{Mixing beliefs among interacting agents}, Advances in Complex Systems
  \textbf{3} (2000), no.~1/4, 87--98.

\bibitem[Gal96]{G96}
Serge Galam, \emph{Fragmentation versus stability in bimodal coalitions},
  Physica A: Statistical and Theoretical Physics \textbf{230} (1996), no.~1-2,
  174 -- 188.

\bibitem[Gal02]{G02}
\bysame, \emph{Minority opinion spreading in random geometry}, European
  Physical Journal B \textbf{25} (2002), 403--406.

\bibitem[Gal05]{G05}
\bysame, \emph{Local dynamics vs. social mechanisms: A unifying frame}, EPL
  (Europhysics Letters) \textbf{70} (2005), no.~6, 705.

\bibitem[Gal08]{G08}
\bysame, \emph{{Sociophysics: A review of Galam models}}, International Journal
  of Modern Physics C \textbf{19} (2008), 409--440.

\bibitem[HD08]{HD08}
S.~Huet and G.~Deffuant, \emph{Bounded confidence with rejection: the infinite
  population limit with perfect uniform initial density}.

\bibitem[HDJ08]{HDJ08}
S.~Huet, G.~Deffuant, and W.~Jager, \emph{A rejection mechanism in 2d bounded
  confidence provides more conformity}, Advances in Complex Systems (ACS)
  \textbf{11} (2008), no.~04, 529--549.

\bibitem[HK02]{HK02}
Rainer Hegselmann and Ulrich Krauze, \emph{Opinion dynamics and bounded
  confidence models, analysis, and simulation}, Journal of Artificial Societies
  and Social Simulation \textbf{5} (2002), no.~3, 8.

\bibitem[InKKB09]{IKKB09}
Gerardo I\~niguez, J\'anos Kert\'esz, Kimmo~K. Kaski, and R.~A. Barrio,
  \emph{Opinion and community formation in coevolving networks}, Phys. Rev. E
  \textbf{80} (2009), no.~6, 066119.

\bibitem[JA05]{JA05}
Wander Jager and Fr\'{e}d\'{e}ric Amblard, \emph{Uniformity, bipolarization and
  pluriformity captured as generic stylized behavior with an agent-based
  simulation model of attitude change}, Comput. Math. Organ. Theory \textbf{10}
  (2005), no.~4, 295--303.

\bibitem[Lat81]{L81}
B.~Latan\'e, \emph{The psychology of social impact}, American Psychologist
  \textbf{36} (1981), 343--356.

\bibitem[Lor07]{L07}
Jan Lorenz, \emph{Continuous opinion dynamics under bounded confidence: A
  survey}, Int. Journal of Modern Physics C \textbf{18} (2007), no.~12,
  1819--1838.

\bibitem[Lor10]{L10}
\bysame, \emph{Heterogeneous bounds of confidence: Meet, discuss and find
  consensus!}, Complexity \textbf{15} (2010), no.~4, 43--52.

\bibitem[RM10]{RM10}
Arezky~H. Rodr\'\i{}guez and Y.~Moreno, \emph{Effects of mass media action on
  the axelrod model with social influence}, Phys. Rev. E \textbf{82} (2010),
  no.~1, 016111.

\bibitem[Sal06]{S06}
Laurent Salzarulo, \emph{A continuous opinion dynamics model based on the
  principle of meta-contrast}, Journal of Artificial Societies and Social
  Simulation \textbf{9} (2006), no.~1, 13.

\bibitem[Sch03]{S03}
Frank Schweitzer, \emph{Brownian agents and active particles: Collective
  dynamics in the natural and social sciences}, Springer Verlag, 2003.

\bibitem[SWS00]{SS00}
Katarzyna Sznajd-Weron and J\'osef Sznajd, \emph{Opinion evolution in closed
  community}, International Journal of Modern Physics C \textbf{11} (2000),
  no.~6, 1157--1165.

\bibitem[VPT10]{VPT10}
{Vaz Martins, T.}, {Pineda, M.}, and {Toral, R.}, \emph{Mass media and
  repulsive interactions in continuous-opinion dynamics}, EPL \textbf{91}
  (2010), no.~4, 48003.

\bibitem[Wei71]{W71}
Wolfgang Weidlich, \emph{The statistical description of polarization phenomena
  in society}, British Journal of Mathematical and Statistical Psychology
  \textbf{24} (1971), 251--266.

\end{thebibliography}

\end{document}